\documentclass{article}%
\usepackage{amsmath}
\usepackage{amsfonts}
\usepackage{amssymb}
\usepackage{graphicx}%
\providecommand{\U}[1]{\protect\rule{.1in}{.1in}}

\begin{document}

\begin{center}
\bigskip\textbf{Features of galactic halo in a brane world model and
observational constraints}

\bigskip

K. K. Nandi$^{1,2,3,a}$, A.I. Filippov$^{3,b}$, F. Rahaman$^{4,c}$,

Saibal Ray$^{5,d}$, A. A. Usmani$^{6,e}$, M. Kalam$^{7,f}$, A.
DeBenedictis$^{8,9,g}$\qquad\ 
\end{center}

$\bigskip$

\begin{center}
$^{1}$Department of Mathematics, University of North Bengal, Siliguri 734 013, India

$^{2}$Joint Research Laboratory, Bashkir State Pedagogical University, Ufa
450000, Russia

$^{3}$Department of Theoretical Physics, Sterlitamak State Pedagogical
Academy, Sterlitamak 453103, Russia

$^{4}$Department of Mathematics, Jadavpur University, Kolkata 700 032, West
Bengal, India

$^{5}$Department of Physics, Government College of Engineering \& Ceramic
Technology, Kolkata 700 010, West Bengal, India

$^{6}$Department of Physics, Aligarh Muslim University, Aligarh 202 002, Uttar
Pradesh, India

$^{7}$Department of Physics, Netaji Nagar College for Women, Regent Estate,
Kolkata 700 092, West Bengal, India

$^{8}$Pacific Institute for the Mathematical Sciences, Simon Fraser University
Site, Burnaby, British Columbia, V5A 1S6, Canada

\qquad$^{9}$Department of Physics, Simon Fraser University, Burnaby, British
Columbia, V5A 1S6, Canada
\end{center}

\bigskip

\begin{center}
$^{a}$E-mail: kamalnandi1952@yahoo.co.in

$^{b}$E-mail: filippovai@rambler.ru

$^{c}$E-mail: farook rahaman@yahoo.com

$^{d}$E-mail: saibal@iucaa.ernet.in

$^{e}$E-mail: anisul@iucaa.ernet.in

$^{f}$E-mail: mehedikalam@yahoo.co.in

$^{g}$E-mail: adebened@sfu.ca

\bigskip

\textbf{ABSTRACT}
\end{center}

Several aspects of the $4d$ imprint of the $5d$ bulk Weyl radiation are
investigated within a recently proposed model solution. It is shown that the
solution has a number of physically interesting properties. The constraints on
the model imposed by combined measurements of rotation curve and lensing are
discussed. A brief comparison with a well known scalar field model is also given.

\bigskip

\textbf{1 INTRODUCTION}

Early observations led to the hypothesis that there could be large amounts of
nonluminous matter hidden in the galactic haloes (Oort 1933; Zwicky 1933,
1937). Later observations of flat rotation curves of spiral galaxies confirmed
the hypothesis (Freeman 1970; Roberts \& Rots 1973; Ostriker, Peebles \&
Yahill 1974; Einasto, Kaasik \& Saar 1974; Rubin, Thonnard \& Ford 1978;
Rubin, Roberts \& Ford 1979; Sofue \& Rubin 2001; Kochanek et al. 2005).
Doppler emissions from stable circular orbits of neutral hydrogen clouds in
the halo allow the measurement of tangential velocity $v_{tg}(r)$ of the
clouds treated as probe particles. According to Newton's laws, centrifugal
acceleration $v_{tg}^{2}/r$ should balance the gravitational attraction
$GM(r)/r^{2}$, which immediately gives $v_{tg}^{2}=GM(r)/r$. That is, one
would expect a fall-off of $v_{tg}^{2}(r)$ with $r$. Observations indicate
that this is not the case:$\ v_{tg}$ approximately levels off with $r$ in the
halo region. The only way to reconcile this result of observation is to
hypothesize that the mass $M(r)$ increases linearly with distance $r$.
Luminous mass distribution in the galaxy does not follow this behavior. Hence
the conclusion that there must be huge amounts of nonluminous matter hidden in
the halo. This unseen matter is given a technical name \textquotedblleft dark
matter". Gravitational lensing measurements have further confirmed the
presence of dark matter (Maoz 1994; Barnes et al. 1999; Cheng \& Krauss 1999;
Sofue \& Rubin 2001; Trott \& Webster 2002; Weinberg \& Kamionkowski 2002;
Kochanek \&\ Schechter 2004; Smith et al. 2005; Faber \& Visser 2006; Metcalf
\& Silk 2007). Current estimates suggest that about 23\% of matter in the
whole universe consists of dark matter residing in the galactic haloes.

Although the exact nature of dark matter is as yet unknown, several candidates
for it have been proposed in the literature. One of the favored candidates is
the Standard Cold Dark Matter of the so called SCDM paradigm (Efstathiou,
Sutherland \& Madox 1990; Pope et al. 2004). Despite its initial success, the
current consensus seems to converge on the Lambda CDM or $\Lambda$CDM model
that is related to the accelerating expansion of the universe (Tegmark et al.
2004 a,b).

Analytic halo models include the framework provided by scalar-tensor theories.
Scalar fields are important because they are predicted by supersymmetric
unification theories. (See for instance, Ellis et al. 1998.) In particular, a
prototype of scalar-tensor theories, namely, the Brans-Dicke theory with
scalar field $\phi$ has the potentiality to explain a wide range of effects,
from those in solar system (Weinberg, 1972; Bhadra, Sarkar \& Nandi 2007) and
gravitational lensing (Bhadra 2003; Nandi, Zhang \& Zakharov, 2006; Sarkar \&
Bhadra 2006) to those arising from objects as exotic as wormholes (Agnese \&
La Camera 1995; Nandi, Evans \&\ Islam 1997; Nandi et al. 1998; Bhadra
\&\ Sarkar 2005). Fay (2004) considered a larger class of scalar-tensor
theories with a potential $V(\phi)$, and with coupling parameter
$\omega=\omega(\phi)$, for the investigation of galactic dynamics. Many other
halo models including suitable variants of scalar field theories also exist in
the literature. See for instance, Bekenstein \& Milgrom (1984), Sanders (1984,
1986), Soleng (1995), Matos, Guzm\'{a}n \& Nu\~{n}ez (2000), Peebles (2000),
Matos \& Guzm\'{a}n (2001), Nucamendi, Salgado \&\ Sudarsky (2001); Mielke \&
Schunck (2002); Cabral-Rosetti et al. (2002), Lidsey, Matos \& Ure\~{n}a-Lopez
(2002), Bharadwaj \& Kar (2003); Arbey, Lesgourgues \&\ Salati (2003).
Substantial halo models are provided by brane world theory (Mak \& Harko 2004;
Rahaman et al. 2008).

The motivation for the brane world model, which we are considering in this
article, comes from an entirely different but important direction. It comes
from the consideration of higher dimensional spacetime because it can be
argued that $4d$ is not big enough to fully accommodate the self interaction
dynamics of gravity. In a recent article, Dadhich (2009) has given persuasive
arguments based on flat space imbedding, self interaction and charge
neutrality. But why only $5d$ and not higher? The reason is that the
conformally flat FRW universe is imbeddable only in $5d$ flat spacetime, which
means that free gravity does not propagate farther than $4d$ because Weyl
curvature vanishes for this FRW universe. The Randall-Sundrum (1999a,b) model
provides the following setting: The $3-$brane (that is, our $4d$ spacetime
which confines matter and other gauge fields) has zero Weyl curvature and it
bounds the $5d$ constant curvature bulk spacetime. To probe into fifth
dimension, one can not rely on light signals as they do not propagate there.
Therefore a possible probe has to rely only on propagation of gravity off the
brane in an unfamiliar non-free manner (Dadhich 2009). For this, one needs to
devise a purely gravity experiment and a method to fathom gravity propagation
in higher dimension. This being a formidable task, the usual approach is to
look for measurable imprints coming from $5d$ into known gravitational
configurations. The galactic halo, where gravity is the king, provides a
natural large scale arena for this observation. It is quite probable that the
effect of unseen dark matter is due to such imprints which contribute to the
source term of Einstein's $4d$ field equations. Solving Einstein's equations,
one then tries to work out the details of higher dimensional non-local effects.

Static spherically symmetric exterior solutions of the brane world model have
been derived in the literature (Dadhich, Maartens, Papadopoulos \& Rezania
2000; Germani \& Maartens 2001; Casadio, Fabbri \& Mazzacurati 2002; Visser \&
Wiltshire 2003; Harko \& Mak 2004; Mak \& Harko 2004; Creek et al. 2006;
Viznyuk \& Shtanov 2007). While all of these works are motivated by one or the
other relevant considerations, the solution by Rahaman et al. (2008) is based
on the simplest supposition of rotation curves that connects dark matter with
the brane world model in a straightforward way. Whatever be the analytic
model, there must be a way to contrast its predictions with actual
measurements. The key point is that one does not directly measure the metric
functions but indirectly measures gravitational potentials and masses from
rotation curve and lensing observations. Bharadwaj \& Kar (2003) advocated
that a combination of rotation curves and lensing measurements could be used
to determine the equation of state of the halo fluid, although their model was
restricted to specific form of flat rotation curve as well as guesses on
equations of state. Faber and Visser (2006) have shown how, in the first post
Newtonian approximation, the combined measurements of rotation curves and
gravitational lensing allow inferences about the mass and pressure profile of
the galactic halo as well as its equation of state.

In this paper, we wish to explore various features of the brane world solution
by Rahaman et al. (2008) focusing in particular on the constraints imposed by
combined measurements. Our general conclusion is that the model has a number
of interesting properties. Nevertheless, it is too premature to say if the
model is observationally supported. The relevant measurement constraints on
the model are worked out. The paper is organized as follows: In this section,
we have already delineated the motivation for the brane world model. In Sec.2,
we briefly outline the field equations and the solution under consideration
for later use. Sec.3 discusses the nature of the galactic fluid while Sec.4
shows stability of circular orbits. In Sec.5, we establish the attractive
nature of gravity in the halo arguing from two different viewpoints and in
Sec.6 we discuss observational constraints. In Sec.7 we make a brief
comparison with the central features of a scalar field model. Sec.8 contains
the conclusions.

\textbf{2 FIELD EQUATIONS AND THE SOLUTION }

The $5d$ bulk field equations with negative vacuum energy $\Lambda_{5}$ can be
reduced to \textquotedblleft effective\textquotedblright\ $4d$ gravitational
field equations on the $3d$ brane. The governing equations are due to
Shiromizu, Maeda \& Sasaki (2000):%
\begin{align}
G_{ij}  &  \equiv R_{ij}-\frac{1}{2}g_{ij}R=-\Lambda g_{ij}+k_{4}^{2}%
T_{ij}+k_{5}^{4}S_{ij}-E_{ij},\\
S_{ij}  &  =\frac{1}{12}TT_{ij}-\frac{1}{4}T_{i}^{q}T_{jq}+\frac{1}{24}%
g_{ij}(3T_{rs}T^{rs}-T^{2}),\nonumber
\end{align}
where the Roman indices run from $0$ to $3$, $k_{5}^{2}=8\pi G_{5}$ is the
$5d$ gravitational constant, $\Lambda$ is the $4d$ cosmological constant given
by $\Lambda=k_{5}^{2}(\Lambda_{5}+k_{5}^{2}\lambda_{b}^{2}/6)$, the $4d$
coupling constant $k_{4}$ is $k_{4}^{2}=k_{5}^{4}\lambda_{b}/6$ and
$\lambda_{b}$ is the vacuum energy on the brane. The stress $E_{ij}$ is the
non-local effect from the bulk given by (Dadhich et al. 2000)
\begin{equation}
E_{ij}=-k^{4}\left[  U\left(  u_{i}u_{j}-\frac{1}{3}h_{ij}\right)
+P_{ij}+2Q_{(i}u_{j)}\right]
\end{equation}
where $k=k_{5}/k_{4}$, $h_{ij}=g_{ij}+u_{i}u_{j}$ is the projection tensor,
$U=k^{4}E_{ij}u^{i}u^{j}$ is the so-called \textquotedblleft dark radiation"
energy density, $Q_{i}$ is a spatial vector while $P_{ij}$ is a spatial, trace
free symmetric tensor. The $E_{ij}$ satisfies, in virtue of the Bianchi
identities, the conservation law%
\begin{equation}
E_{j;i}^{i}=k_{5}^{4}S_{j;i}^{i}%
\end{equation}
where semicolon denotes covariant derivative with respect to $g_{ij}$. The
simplest case is the case of vacuum (absence of ordinary matter) so that
$T_{ij}=0\Rightarrow S_{ij}=0$ which, in the paradigm of the study here,
implies that we are interested in the region away from the galactic core,
where the dark matter strongly dominates over the ordinary matter. In static
vacuum, we have $Q_{i}=0$ and in the comoving orthonormal frame, we have
$u^{i}=\delta_{0}^{i}$, $h_{ij}=$diag$(0,1,1,1)$. In this case, the constraint
Eq (3) suggests a solution (Dadhich et al.\textit{ }2000)%
\begin{equation}
P_{ij}=P[r_{i}r_{j}-\frac{1}{3}h_{ij}]
\end{equation}
where $r_{i}$ is a unit radial vector and $P$ is the so-called
\textquotedblleft dark pressure"$.$

Let us consider a spherically symmetric line element
\begin{equation}
ds^{2}=-e^{\nu(r)}dt^{2}+e^{\lambda(r)}dr^{2}+r^{2}(d\theta^{2}+sin^{2}\theta
d\phi^{2})
\end{equation}
where $\nu$ and $\lambda$ are the metric potentials and are function of the
space coordinate $r$ only, such that $\nu=\nu(r)$ and $\lambda=\lambda(r)$.
For this symmetry, $P_{ij}=P(r)[r_{i}r_{j}-\frac{1}{3}h_{ij}]$. In static
vacuum ($T_{ij}=0,Q_{i}=0$) with the trace free source term $E_{i}^{i}=0$
$\Rightarrow$ $R_{i}^{i}=0$, and with the unit radial vector $r_{i}%
=(0,1,0,0)$, the field equations (1) yield the following (Mak \& Harko 2004):
\begin{equation}
e^{-\lambda}\left(  \frac{\lambda^{\prime}}{r}-\frac{1}{r^{2}}\right)
+\frac{1}{r^{2}}=\frac{48\pi}{k^{4}{\lambda}_{b}}U,
\end{equation}%
\begin{equation}
e^{-\lambda}\left(  \frac{\nu^{\prime}}{r}+\frac{1}{r^{2}}\right)  -\frac
{1}{r^{2}}=\frac{16\pi}{k^{4}{\lambda}_{b}}(U+2P),
\end{equation}%
\begin{equation}
\frac{e^{-\lambda}}{2}\left(  \nu^{{\prime}{\prime}}+\frac{{\nu^{\prime}}^{2}%
}{2}-\frac{{\nu^{\prime}\lambda^{\prime}}}{2}+\frac{\nu^{\prime}%
-\lambda^{\prime}}{r}\right)  =\frac{16\pi}{k^{4}{\lambda}_{b}}(U-P),
\end{equation}%
\begin{equation}
\nu^{\prime}=-\frac{1}{2U+P}\left[  U^{\prime}+2P^{\prime}+\frac{6P}%
{r}\right]  ,
\end{equation}
where $U=U(r)$, $P=P(r)$ and primes denote differentiation with respect to
$r$. [\textit{Eq.(7)} of Rahaman et al. (2008) has been properly rearranged in
Eq.(8) above, which reveals the correct Ricci component on the left hand side].

Assuming the known flat rotation curve condition $v_{tg}^{2}=\frac{r(e^{\nu
})^{\prime}}{2e^{\nu}}=$ constant (Chandrasekhar 1983), Rahaman et al.\textit{
}(2008) obtained a solution as follows:%
\begin{equation}
e^{\nu(r)}\equiv B(r)=B_{0}r^{l},
\end{equation}%
\begin{equation}
e^{\lambda(r)}\equiv A(r)=\left[  \frac{2}{(2+\frac{l}{2})a}+\frac{D}{r^{a}%
}\right]  ^{-1},
\end{equation}
where $l=2(v_{tg})^{2}$, $B_{0}>0$ and $D$ are arbitrary constants and
$a=(2+l+\frac{l^{2}}{2})/(2+\frac{l}{2})$. They showed that the mass function
$M(r)=\frac{48\pi}{k^{4}{\lambda}_{b}}\int Ur^{2}dr$ increases linearly with
$r$, which agrees with observation (Begeman 1989). However, this expression
for mass is Newtonian and is applicable only so long as one can neglect
pressure contributions although as yet there is no observational ground for
such a neglect. To see the expressions for various types of mass distributions
in terms of observational parameters, we wait till Sec.6. Meanwhile, there are
several other facets of the solution that need attention, which are addressed below.

\textbf{3 NATURE OF DARK RADIATION }

To throw some light on this issue, we identify the right hand side of the
equations (6)-(8) with the standard spherically symmetric fluid distribution
($\rho$, $p_{r}$, $p_{t}$, $p_{t}$) via Einstein's equations $G_{\widehat
{i}\widehat{j}}=-8\pi E_{\widehat{i}\widehat{j}}$ in the observer's
orthonormal rest frame ($\symbol{94}$). Using Eqs.(10) and (11) in (6)-(8), we
get:
\begin{align}
G_{\widehat{t}\widehat{t}}  &  =U=\rho=\left(  \frac{1}{8\pi}\right)  \left[
\frac{l(l+2)r^{-2}}{\beta}+\frac{Dr^{-\alpha}l(l+1)}{l+4}\right] \\
G_{\widehat{r}\widehat{r}}  &  =\frac{U+2P}{3}=p_{r}=\left(  \frac{1}{8\pi
}\right)  \left[  \frac{Dr^{-\alpha}(l^{3}+3l^{2}+6l+4)-(l-2)r^{-2}}{\beta
}\right] \\
G_{\widehat{\theta}\widehat{\theta}}  &  =G_{\widehat{\varphi}\widehat
{\varphi}}=\frac{U-P}{3}=p_{t}=\left(  \frac{1}{8\pi}\right)  \left[
\frac{l^{2}(l+4)r^{-2}-2Dr^{-\alpha}(l^{3}+3l^{2}+6l+4)}{l^{3}+6l^{2}%
+12l+16}\right] \\
\alpha &  \equiv\frac{l^{2}+4l+12}{l+4},\text{ \ \ }\beta\equiv l^{2}+2l+4
\end{align}
where we have chosen units on the right hand side of Eqs.(6)-(8) such that
$\frac{48\pi}{k^{4}{\lambda}_{b}}=1$ and redefined the fluid stresses ($\rho$,
$p_{r}$, $p_{t}$, $p_{t}$) in terms of $U$ and $P$. Eliminating $U$ from
Eqs.(13), (14), one can easily find the expression for $P(r)$ as given in
Rahaman et al. (2008) after their arbitrary constant of integration is set to
zero. There is actually no room for a nonzero constant because once the metric
functions are known, $U$ and $P$ follow from Eqs. (6)-(8).

The interesting results we find are the following: While $U$ is the same as
$\rho$, the dark pressure $P(r)$ is equal to neither $p_{r}$ nor $p_{t\text{
}}$but $P(r)=p_{r}(r)-p_{t}(r)\neq0$. This indicates that the dark Weyl
radiation imprint can not be characterized in the rest frame by a perfect
fluid which requires $p_{r}(r)=p_{t}(r)$. So the perfect fluid version of dark
radiation is that $U(r)\neq0$ but $P(r)=0$. Looking at Eq.(2), we see that the
imprint radiation in $4d$ vacuum is a dust-like fluid where $E_{ij}%
=-k^{4}U(r)(u_{i}u_{j}-\frac{1}{3}h_{ij})$. This is not the case with the
solution under consideration since $p_{r}(r)\neq p_{t}(r)$. This pressure
anisotropy is a good feature of the solution from the point of view of
exterior matching. Note that the solution cannot be matched to the
Schwarzschild exterior metric at the boundary of the halo if the pressures
were isotropic (Bharadwaj \& Kar 2003). The Schwarzschild solution, although
not the unique spherically symmetric vacuum when considering the brane-world
model, is desirable for mimicking the far region (vacuum) of true $4d$
gravity, as supported by observation. We further see that the dark radiation
fluid is not of exotic nature. The minimal condition for exoticity is that the
Null Energy Condition (NEC) should be violated (Hochberg \&Visser 1998). We
find that the NEC is satisfied at all radii because $\rho+p_{r}\geq0$ (Fig.1).
Transverse pressures are not included as they refer only to normal matter
(Visser, Kar \& Dadhich 2003). Even if we include them, we find $\rho
+p_{r}+2p_{t}\geq0$ for all $r$.

In the rest of the article, we assume the following numerical values for
constants. Observations of the frequency shifts in the HI radiation show that,
in the halo region, $v_{tg}/c$ is nearly constant at a value $7\times10^{-4}%
$~(Binney \&\ Tremaine 1987; Persic, Salucci \& Stel 1996; Boriello \& Salucci
2001). Thus $l\sim10^{-6}$. We also note that the solution is not used for
small $r$, the inner core of the galaxy. Accordingly, to be consistent with
observational facts, in all the figures in this paper we shall take $B_{0}=1$,
$l=0.000001$, $D=0.00001$ and fairly large distances in $Kpcs$.

\textbf{4 STABILITY OF CIRCULAR ORBITS}

Defining the four velocity $U^{\alpha}=\frac{dx^{\sigma}}{d\tau}$ of a test
particle moving solely in the subspace of the brane (and restricting ourselves
to $\theta=\pi/2$), the equation $g_{\nu\sigma}U^{\nu}U^{\sigma}=-m_{0}^{2}$
can be cast in a Newtonian form
\begin{equation}
\left(  \frac{dr}{d\tau}\right)  ^{2}=E^{2}+V(r)
\end{equation}
which gives%
\begin{equation}
V(r)=-\left[  E^{2}\left\{  1-\frac{r^{-l}A^{-1}}{B_{0}}\right\}
+A^{-1}\left(  1+\frac{L^{2}}{r^{2}}\right)  \right]
\end{equation}%
\begin{equation}
E=\frac{U_{0}}{m_{0}},L=\frac{U_{3}}{m_{0}},
\end{equation}
where the constants $E$ and $L$, respectively, are the conserved relativistic
energy and angular momentum per unit rest mass of the test particle. Circular
orbits are defined by $r=R=$constant so that $\frac{dR}{d\tau}=0$ and,
additionally, $\frac{dV}{dr}\mid_{r=R}=0$. From these two conditions follow
the conserved parameters:
\begin{equation}
L=\pm\sqrt{\frac{l}{2-l}}R
\end{equation}
and using it in $V(R)=-E^{2}$, we get
\begin{equation}
E=\pm\sqrt{\frac{2B_{0}}{2-l}}R^{\frac{l}{2}}.
\end{equation}
The orbits will be stable if $\frac{d^{2}V}{dr^{2}}\mid_{r=R}<0$ and unstable
if $\frac{d^{2}V}{dr^{2}}\mid_{r=R}>0$. Putting the expressions for $L$ and
$E$ in $\frac{d^{2}V}{dr^{2}}\mid_{r=R}$, we obtain, after straightforward
calculations, the final result, viz.,
\begin{equation}
\frac{d^{2}V}{dr^{2}}\mid_{r=R}=-\frac{2lR^{-2-a}\{4R^{a}+(4+2l+l^{2}%
)D\}}{4+2l+l^{2}}.
\end{equation}
Thus $\frac{d^{2}V}{dr^{2}}\mid_{r=R}<0$ so that circular orbits are stable in
the model under consideration when $D>0$. The latter condition is always
satisfied since, from Eq. (11), we see that $D$ has a dimension proportional
to a power of radius $R$. The radius is always positive and so is $D$.

\textbf{5 ATTRACTION IN DARK RADIATION}

Now let us return to the question of attractive gravity. Observations indicate
that gravity on the galactic scale is attractive (clustering, structure
formation etc). By the existence of stable circular orbits we already know
that the particles are being accelerated towards the galactic center. This can
also be seen by studying the geodesic equation for a test particle that has
been \textquotedblleft placed\textquotedblright\ at some radius $r_{0}$, which
yields
\begin{equation}
\frac{d^{2}r(\tau)}{d\tau^{2}}|_{r_{0}}=-\frac{B_{0}l}{2\,r^{1-l}}\left[
\frac{2}{\left(  2+\frac{l}{2}\right)  a}+\frac{D}{r^{a}}\right]  \left(
\frac{dt(\tau)}{d\tau}\right)  ^{2}|_{r_{0}}\,.
\end{equation}

The quantity in square brackets is $g^{rr}$ and therefore must be positive.
Therefore this expression is negative and it can be deduced that particles are
attracted towards the center. Detailed discussions on this quantity may be
found in Ford \& Roman ~(1996) and Lobo~(2008). A related issue is the
following. In the Newtonian limit the gravitational energy of a spherically
symmetric, gravitationally attractive system, with sufficient fall-off
properties is negative. One could potentially argue that the $4d$
gravitational effects due to higher dimensional Weyl stresses on the brane
should not be subject to the same criteria as in true four-dimensional
gravity. However, from the point of view of the Newtonian limit on the $3d$
brane, as well as CMB studies, it is desirable that the contribution from the
Weyl stresses not be too large away from the galaxy. The question then becomes
one of how to quantify such an effect in the relativistic case.

We choose as our quantifier the following: The total gravitational energy
$E_{G}$ in the halo region must be negative~(Misner, Thorne \& Wheeler 1973;
Lynden-Bell, Katz \& Bi\v{c}\'{a}k 2007), where by gravitational energy we
mean the quantity defined below. It is necessary to explicitly verify it
because a positive energy density does not always lead to attractive
gravity~(Nandi et al. 2009). Hence we shall calculate the total gravitational
energy $E_{G}$ between two arbitrary but fixed radii with $r>0$ (we have
excluded the origin since the solution is not valid at small $r$ because of
singularity there, and the effects of the core region are manifest in the
constants in the solution). This quantity is given by~(Misner, Thorne \&
Wheeler 1973; Lynden-Bell, Katz \& Bi\v{c}\'{a}k 2007):
\begin{equation}
E_{G}=M-E_{M}=4\pi\int_{r_{1}}^{r_{2}}[1-A^{\frac{1}{2}}]\rho r^{2}dr
\end{equation}
where $1-A^{\frac{1}{2}}<0$ by definition (proper radial length is larger than
the Euclidean length) and%
\begin{equation}
M=4\pi\int_{r_{1}}^{r_{2}}\rho r^{2}dr,
\end{equation}
$E_{M}$ is the sum of other forms of energy like rest energy, kinetic energy,
internal energy etc. The dark radiation energy density (12) can be rewritten
as~
\begin{equation}
\rho=\frac{1}{8\pi}\left[  \frac{D(a-1)}{r^{a+2}}+\left(  1-\frac{2}%
{a(2+\frac{l}{2})}\right)  \frac{1}{r^{2}}\right]  .
\end{equation}
For very small values of $D$ and for $0<l\ll2$, it follows from the
integration in Eq.(24) that $E_{G}<0$ for arbitrary $r_{2}>r_{1}>0$,
indicating that gravity in the halo is indeed attractive. It is observed that
the galactic halo extends as far as $500Kpc$~(Sahni 2004). Thus as an
illustration we might take in units of $Kpc$, $r_{1}=100$, $l=10^{-6}$, then
it follows from Fig.2 that $E_{G}<0$ for the upper limit $r_{2}<500$. We
recommend a value of $D\leq0.00001$, because it optimally yields reasonable
behavior for all relevant quantities in the entire range of the halo

\textbf{6 OBSERVATIONAL CONSTRAINTS }

We wish to contrast the brane world model with the observational parameters in
the galactic halo. For the ease of calculation, we shall rewrite the metric
(5)\ in the form%
\begin{equation}
ds^{2}=-e^{2\Phi(r)}dt^{2}+\frac{dr^{2}}{1-\frac{2m(r)}{r}}+r^{2}(d\theta
^{2}+\sin^{2}\theta d\varphi^{2})
\end{equation}
where, for the solution under consideration, the metric functions are
specifically given by
\begin{equation}
\Phi(r)=\frac{1}{2}[\ln B_{0}+l\ln r]
\end{equation}%
\begin{equation}
m(r)=\frac{r}{2}[1-A^{-1}(r)].
\end{equation}
These functions are not necessarily the same as the potential and mass
functions obtained from observations. Note from equation (28) that $A>1$,
which is a basic condition to be satisfied by a valid solution. In the first
post Newtonian approximation, the gravitational potential $\Phi(r)$ is given
in general relativity by the equation (Misner, Thorne \& Wheeler 1973):%
\begin{equation}
\nabla^{2}\Phi\approx R_{tt}\approx4\pi(\rho+p_{r}+2p_{t})
\end{equation}
which reduces to the equation for Newtonian $\Phi_{N}$ where%
\begin{equation}
\nabla^{2}\Phi_{N}=4\pi\rho
\end{equation}
only if the pressure contributions are negligible in comparison to energy density.

The above general relativistic $\Phi$ appears in the wavelength shifts
$z_{\pm}$ of an emission line of a massive probe particle for an edge-on
galaxy \ is (Nucamendi, Salgado \&\ Sudarsky 2001; Lake 2004)
\begin{equation}
1+z_{\pm}=\frac{1}{\sqrt{1-r\Phi^{\prime}(r)}}\left[  \frac{1}{e^{\Phi(r)}%
}-\frac{\pm\left\vert b\right\vert \sqrt{r\Phi^{\prime}(r)}}{r}\right]
\end{equation}
where $b$ is the impact parameter and $\pm$ signs refer to approaching and
receding particles. The above equation approximates to (Faber \& Visser 2006)
\begin{equation}
z_{\pm}^{2}\approx r\Phi^{\prime}(r).
\end{equation}
Therefore, the usual techniques for obtaining the potential for rotation curve
measurements yield a \textquotedblleft pseudo-potential"%
\begin{equation}
\Phi_{\text{RC}}=\Phi\neq\Phi_{N}.
\end{equation}
From Eq.(29), one then obtains a \textquotedblleft pseudo-mass" as
\begin{equation}
m_{\text{RC}}=r^{2}\Phi^{\prime}(r)\approx4\pi\int(\rho+p_{r}+2p_{t})r^{2}dr.
\end{equation}
The right hand side expression comes from the first post Newtonian order in
the weak field limit and we call it $M_{pN}$. We have so distinguished it
because $r^{2}\Phi^{\prime}(r)$ comes entirely from pseudo potential $\Phi$
whereas $M_{pN}$ is an integrated quantity that requires the knowledge of the
equation of state. The pseudo mass $m_{\text{RC}}$ reduces to the Newtonian
mass $M(r)$ [see Eq.(24)] when pressure contributions are negligible in
comparison to energy density.

Photons can also be regarded as probe particles as they sense the
gravitational field in the halo during their travel to the observer. The
effect of gravity on photon motion can be calculated in terms of a refractive
index $n(r)$. (See also Boonserm et al. 2005.) For the Schwarzschild gravity,
the exact geodesic equation, including the equation for Shapiro time delay, of
massless particles was expressed in terms of $n(r)$ by Nandi \& Islam (1995)
via the idea of optical-mechanical analogy. This idea has been further
extended by Evans et al\textit{. }(2001) to \textit{arbitrary} spherically
symmetric spacetimes and by Alsing (1998) to rotating spacetimes. The geodesic
motion of both massless and massive particles could be exactly expressed in
terms of a single generalized refractive index $N=n^{2}V/c$ where $V$ is the
$3-$velocity of the particle. For light motion, $V=c/n$.

For unspecified metric functions $\Phi(r)$ and $m(r)$, Faber and Visser (2006)
argue that $n(r)=1-2\Phi_{\text{lens}}+O[\Phi_{\text{lens}}^{2}]$ in which
they define the lensing pseudo-potential as
\begin{equation}
\Phi_{\text{lens}}=\frac{\Phi(r)}{2}+\frac{1}{2}\int\frac{m(r)}{r^{2}}dr.
\end{equation}
Another pseudo-mass $m_{\text{lens}}$ obtained from lensing measurements has
been defined as (Faber \& Visser 2006)%

\begin{equation}
m_{\text{lens}}=\frac{1}{2}r^{2}\Phi^{\prime}(r)+\frac{1}{2}m(r).
\end{equation}
The first order approximations of Einstein's equations yield%
\begin{equation}
\rho(r)\approx\frac{1}{4\pi r^{2}}[2m_{\text{lens}}^{\prime}(r)-m_{\text{RC}%
}^{\prime}(r)]
\end{equation}%
\begin{equation}
4\pi r^{2}(p_{r}+2p_{t})\approx2[m_{\text{RC}}^{\prime}(r)-m_{\text{lens}%
}^{\prime}(r)]
\end{equation}
where the right hand sides denote pseudo-density and pseudo-pressures.
Furthermore, Faber \& Visser (2006) defined a dimensionless quantity%
\begin{equation}
\omega(r)=\frac{p_{r}+2p_{t}}{3\rho}\approx\frac{2}{3}\frac{m_{\text{RC}%
}^{\prime}-m_{\text{lens}}^{\prime}}{2m_{\text{lens}}^{\prime}-m_{\text{RC}%
}^{\prime}}%
\end{equation}
where the pseudo quantities on the right hand side of Eqs.(37)-(39) determine,
respectively, the observed density, pressure and equation of state. Using the
specific metric functions (27, 28) in Eqs.(35)-(39), we can calculate the
first order quantities $\Phi_{\text{RC}}$, $m_{\text{RC}}$, $\Phi
_{\text{lens}}$, $m_{\text{lens}}$, $2(m_{\text{RC}}^{\prime}-m_{\text{lens}%
}^{\prime})$ and $\frac{2}{3}\frac{m_{\text{RC}}^{\prime}-m_{\text{lens}%
}^{\prime}}{2m_{\text{lens}}^{\prime}-m_{\text{RC}}^{\prime}}$ to be expected
from combined observations.

Note that pseudo quantities are the observable quantities. The general
procedure to determine the metric can be stated as follows: Once one is able
to observationally determine the profiles of pseudo quantities, one can work
backwards to find the corresponding metric functions. This is a kind of
reverse technique observational astrophysicists use. If the observed pseudo
profiles fit (up to experimental error) with the analytic profiles of
\textit{a priori} given metric functions, one can say that the solution is
physically substantiated. Otherwise, it has to be ruled out as non-viable. In
this sense, the observed pseudo profiles play the role of constraints on the
possible metric solutions.

For the present solution, we obtain the following constraint equations on
$\Phi(r)$, $m(r)$, the pressure profile and the equation of state:%

\begin{equation}
\Phi_{\text{RC}}=\frac{1}{2}(\ln B_{0}+l\ln r)
\end{equation}%
\begin{equation}
m_{\text{RC}}=r^{2}\Phi^{\prime}(r)=\frac{lr}{2}%
\end{equation}%
\[
\Phi_{\text{lens}}=\frac{1}{4}\left[  (\ln B_{0}+l\ln r)+\frac{D(l+4)r^{-\frac
{\beta}{l+4}}+l(l+2)\ln r}{\beta}\right]
\]%
\begin{equation}
m_{\text{lens}}=\frac{1}{4}\left[  \frac{l(l^{2}+3l+6)r}{\beta}-Dr^{-\gamma
}\right]
\end{equation}%
\begin{equation}
2(m_{\text{RC}}^{\prime}-m_{\text{lens}}^{\prime})=l-\frac{1}{2}\left[
\frac{l(l^{2}+3l+6)}{\beta}+\gamma Dr^{-(1+\gamma)}\right]
\end{equation}

\begin{align}
\omega(r)  &  =\frac{2}{3}\frac{m_{\text{RC}}^{\prime}-m_{\text{lens}}%
^{\prime}}{2m_{\text{lens}}^{\prime}-m_{\text{RC}}^{\prime}}=\frac{1}{3}%
\frac{(l^{3}+5l^{2}+6l+8)r^{1+\gamma}-D(l^{3}+3l^{2}+6l+4)}{D(l^{3}%
+3l^{2}+6l+4)+(l^{2}+6l+8)r^{1+\gamma}}\\
\gamma &  \equiv\frac{l(l+1)}{l+4}.\nonumber
\end{align}

If observations yield the same profiles as on the right hand sides, we can say
that they support the metric functions $\Phi(r)$ and $m(r)$ of Eqs.(27, 28).
Assuming for the moment that these specific functions are really
observationally valid, we ask what comparative features of the observables do
we expect to see. These can be visualized by first comparing the functions
$\Phi_{\text{RC}}$ with $\Phi_{\text{lens}}$. We plot their difference in
Fig.3 taking the numerical values of constants mentioned earlier and observe
that despite their varying expressions, the actual difference $\Phi
_{\text{RC}}-\Phi_{\text{lens}}$ is negligible, of the order of only
$\sim10^{-6}$. Thus rotation curve and lensing measurements lead to
practically indistinguishable inferences about the pseudo potential, at least
in the first order. Similarly, Fig.4 shows variation with distance of
$M_{pN}(r)$ and of different pseudo-masses $m_{\text{RC}}$ and $m_{\text{lens}%
}$. All the masses share the common feature that they increase with distance
$r$, though they begin to slightly differ in the far field. To ascertain the
role of pressures from the observational standpoint, one has to look at Fig.5
where we find that the difference $m_{\text{RC}}^{\prime}-m_{\text{lens}%
}^{\prime}$ is of order $\sim10^{-7}$, to first order. To ascertain the
equation of state from observational parameters, we come to Eq.(44), which is
plotted in Fig.6. One immediately finds that $\omega\rightarrow1/3$. The
closeness to $1/3$ actually depends on the value of $D$, the smaller its value
the better the closeness. The exact equation of state following directly from
the Eqs.(12)-(14) is
\begin{equation}
p_{r}+2p_{t}=\rho
\end{equation}
which implies from Eq.(39) that
\begin{equation}
\omega=\frac{1}{3}.
\end{equation}
(Had the dark radiation been a perfect fluid, we could have $p=\frac{1}{3}%
\rho$, the usual equation of state for radiation.) Noting that $U=\rho$ and
using Eq.(45) in Eq.(34) we find that the mass in the first post Newtonian
approximation becomes%
\begin{equation}
M_{pN}(r)=4\pi\int(\rho+p_{r}+2p_{t})r^{2}dr=2M(r)
\end{equation}
which is twice the Newtonian mass $M(r)$ defined by Eq.(24). All the above
indicate that the galactic halo modelled by the present solution shows
somewhat, but not quite, Newtonian features (pressure contribution is not
negligible as it is proportional to energy density). The profiles of (pseudo)
potentials, masses and the anisotropic equation of state as expressed in
Eqs.(40)-(44) above are the ones one would expect to find from the combined
measurement of the pseudo quantities, \textit{if} the solution (27, 28) is
really a viable one.

\begin{center}
\textbf{7 COMPARISON WITH A SCALAR FIELD MODEL}
\end{center}

Scalar fields are an integral part of reality though they yet lack any
observational evidence. We choose the well discussed scalar field model with
potential $\widetilde{V}(r)$ proposed by Matos, Guzm\'{a}n \& Nu\~{n}ez (2000)
and compare it with the present brane world model. Using the flat rotation
curve condition, they obtain the full solution as (we distinguish their
quantities by tilde):%
\begin{equation}
\widetilde{B}(r)=B_{0}r^{l}%
\end{equation}%
\begin{equation}
\widetilde{A}(r)=\frac{4-l^{2}}{4+D(4-l^{2})r^{-(l+2)}}%
\end{equation}%
\begin{equation}
\widetilde{\phi}(r)=\sqrt{\frac{l}{8\pi}}\ln r+\phi_{0}%
\end{equation}%
\begin{equation}
\widetilde{V}(r)=-\frac{1}{8\pi(2-l)r^{2}},
\end{equation}
where $D$ is an arbitrary constant of integration. For simplicity, they take
$D=0$, but this assumption makes $\widetilde{A}<1$ [see equation (28)].
Because of this, their expressions for density and pressures pressures lead to
different conclusions than those in the brane-world model considered here. For
example, the expression for density exhibit $\widetilde{\rho}<0$, meaning
violation of Weak Energy Condition (WEC) and furthermore it leads to
$\widetilde{\omega}<-1$, meaning repulsive gravity in the halo, contradicting
observational facts. Therefore it is necessary to re-calculate the relevant
quantities with $D\neq0$.

We find the density and pressure profiles in the rest frame of the fluid as%
\begin{equation}
\widetilde{\rho}=\frac{1}{8\pi}\frac{r^{-(4+l)}[D(l^{3}+l^{2}-4l-4)+l^{2}%
r^{2+l}]}{l^{2}-4}%
\end{equation}%
\begin{equation}
\text{ }\widetilde{p}_{r}=\frac{1}{8\pi}\frac{r^{-(4+l)}[D(l^{3}%
+l^{2}-4l-4)-l(4+l)r^{2+l}]}{l^{2}-4}%
\end{equation}%
\begin{equation}
\text{ }\widetilde{p}_{t}=\frac{1}{8\pi}\frac{r^{-(6+l)}[D(l^{3}%
+l^{2}-4l-4)+l^{2}r^{2+l}][(r^{2}-1)l-2(r^{2}+1)]}{4(l^{2}-4)}.
\end{equation}
We wish to emphasize here that the role of non-zero value of $D$ is crucial
not only for avoiding repulsive gravity but also for arriving at a correct
conclusion about the relative strengths between pressure and density. For
instance, let us take $D=1$. In the distant halo region, we can take, say,
$r\sim100-300$ $Kpc$ and with $l\sim10^{-6}$, we find the numerical values to
be $\widetilde{\rho}\sim10^{-9\text{ }}$ and $\widetilde{p}_{r}\sim10^{-9}$,
which means that they are of the same order. But $\widetilde{p}_{r}%
+2\widetilde{p}_{t}\sim10^{-11}\Rightarrow\widetilde{p}_{r}+2\widetilde{p}%
_{t}\sim10^{-2}\widetilde{\rho}$, which indicates that total pressure is
roughly one hundred times less than the density. All these clearly go against
the conclusion of the authors. However, if we take $D=0.00001$, we find that
$\widetilde{p}_{r}+2\widetilde{p}_{t}\sim10^{3}\widetilde{\rho}$. If we keep
on decreasing the value of $D$ further (but never exactly to zero for reasons
stated above) we see that the total pressure dominates more and more over
density so that the system becomes indeed non-Newtonian as claimed by Matos,
Guzm\'{a}n \& Nu\~{n}ez (2000). Such minuscule values of $D$ are remarkably
similar to those recommended for our solution (see Sec.5).

The next question is how far can we decrease $D$? We notice the following
interesting scenario:\ When $D=10^{-7}$, we find $\widetilde{p}_{r}%
+2\widetilde{p}_{t}=9\times10^{5}\widetilde{\rho}$, which leads to
$\widetilde{\omega}=\frac{\widetilde{p}_{r}+2\widetilde{p}_{t}}{3\widetilde
{\rho}}=3\times10^{5}$. This is the extreme non-Newtonian model possible in
the scalar field model. On the other hand, if $D=10^{-8}$, we find that
$\widetilde{\omega}>0$ up to $r=r_{0}=200$ $Kpc$ (attractive gravity) and
becomes $\widetilde{\omega}<-1$ after $r=r_{0}$ (repulsive gravity). At
$r=r_{0}$, there is a singularity in $\widetilde{\omega}$. When $D=10^{-9}$,
we find that $\widetilde{\rho}<0$, $\widetilde{\omega}<-1$, exhibiting the
characteristics that follow from the same choice $D=0$. Therefore we conclude
that the limiting value of $D$ is $10^{-7}$.

The equation of state is anisotropic, as is evident from equations (54)-(55),
a feature shared also by the brane world solution. It can be verified that
$\widetilde{\rho}>0$, $\widetilde{\rho}+\widetilde{p}_{r}>0$, $\widetilde
{\rho}+\widetilde{p}_{r}+2\widetilde{p}_{t}>0$ for $D\geq10^{-7}$, so we can
say that the halo matter is not exotic because the WEC and NEC are satisfied
everywhere. Therefore, we expect an attractive halo. To confirm it, again we
follow the prescription in Lynden-Bell, Katz \& Bi\v{c}\'{a}k (2007), and find
that the total gravitational energy is indeed negative:
\begin{equation}
\widetilde{E}_{G}=4\pi\int_{r_{1}}^{r_{2}}[1-\widetilde{A}^{\frac{1}{2}%
}]\widetilde{\rho}r^{2}dr<0,
\end{equation}
due to the fact that $\widetilde{\rho}>0$, $1-\widetilde{A}^{\frac{1}{2}}<0$
and $r_{2}>r_{1}$.

Certainly, the extreme scalar field model corresponding to $D=10^{-7}$ is
highly non-Newtonian, that is, $\widetilde{p}_{r}+2\widetilde{p}_{t}\sim
10^{6}\widetilde{\rho}$. As a result, Eq.(30) leading to a purely Newtonian
definition of mass $M(r)$ as in Eq.(24) does not apply. However, incorporating
the pressure contribution, the dynamical mass in the first post Newtonian
order is
\begin{equation}
\widetilde{M}_{pN}(r)=4\pi\int(\widetilde{\rho}+\widetilde{p}_{r}%
+2\widetilde{p}_{t})r^{2}dr=10^{6}\widetilde{M}(r),
\end{equation}
which clearly reflects the non-Newtonian nature of the model in terms of
masses. The observable pseudo quantities for this extreme case work out to
\begin{equation}
\widetilde{m}_{\text{RC}}(r)=\frac{lr}{2}\approx10^{-6}r
\end{equation}%
\begin{equation}
\widetilde{m}_{\text{lens}}(r)\approx\frac{l(l^{2}+l-4)r}{4(l^{2}-4)}%
\approx10^{-6}r
\end{equation}%
\begin{equation}
2(\widetilde{m}_{\text{RC}}^{\prime}-\widetilde{m}_{\text{lens}}^{\prime
})\approx\frac{l(l^{2}-l-4)}{2(l^{2}-4)}\approx10^{-6}%
\end{equation}%
\begin{equation}
\widetilde{\omega}(r)\approx\frac{2}{3}\frac{\widetilde{m}_{\text{RC}}%
^{\prime}-\widetilde{m}_{\text{lens}}^{\prime}}{2\widetilde{m}_{\text{lens}%
}^{\prime}-\widetilde{m}_{\text{RC}}^{\prime}}=\frac{l(l^{2}-l-4)r^{2+l}%
-D(l^{3}+l^{2}-4l-4)}{3[D(l^{3}+l^{2}-4l-4)+l^{2}r^{2+l}]}\approx3\times10^{5}%
\end{equation}
for our chosen range, $r\sim100-300$ $Kpc$. Note that if we straightaway put
$D=0$ in equation (60), we get $\widetilde{\omega}(r)<-1$, conveying a
completely different conclusion. Thus while the numerical values of observable
potentials and masses behave like those in the brane solution, it is only the
equation of state (61) that differs widely because $\widetilde{\omega}$ has a
value $\sim10^{5}$ compared to $\omega=\frac{1}{3}$ [cf.Eq.(47)].

\textbf{8 CONCLUSIONS }

In the foregoing, we first discussed the motivation for considering higher
dimensions, which probably distinguishes the solution (27,28) as a more
interesting search for the halo model. It is the $3-$brane that we experience
and therefore the bulk contributions $U$ and $P$ are translated into brane
quantities via Einstein equations, which provide a way to have insights into
the nature of the galactic fluid. The present model shows that the halo does
not have a perfect fluid equation of state but NEC is preserved, meaning that
the fluid is not exotic. It was shown that circular orbits in the halo are
stable and that the dark radiation is attractive in nature, as it must be.
Thus the solution satisfies two crucial physical requirements: Circular orbit
stability and attractive gravity in the halo. The next task was to derive the
constraints on the solution imposed by combined observations of rotation curve
and lensing.

The analyses by Nucamendi, Salgado \& Sudarsky (2001), Lake (2004) and Faber
\&\ Visser (2006) have been carried out without reference to any specific form
of metric functions. The specific functions (27,28) yielded the right hand
sides of Eqs.(40)-(44) in terms of parameters to be measured by a combination
of rotation curve and lensing measurements. These are effectively the
constraint equations on the chosen solution. If combined observations turn out
to tally with the behavior predicted by Eqs.(40)-(44), then the brane world
solution (27, 28) can be said to be supported by observation. Until that
happens, the solution would remain largely an academic curiosity.

As discussed in the introduction, the increase of mass linearly with $r$ comes
from an elementary Newtonian argument to explain the observed fact of flat
rotation curves. A key question still remains: How much of pressure
contribution is there in the making of that mass, that is, is it just the
Newtonian $M(r)$ of Eq.(24) or $M_{pN}(r)$ of Eq.(34)? The solution by Rahaman
et al. (2008) does exhibit a linear increase of the Newtonian mass $M(r)$.
Fortunately, even if pressures are included, we find $M_{pN}(r)=2M(r)$
[cf.Eq.(47)], that is, they are of the same order showing that the theoretical
linear mass increase is pretty consistent with both the definitions within the
present model. Such an increase is also supported by observable pseudo masses,
as expected. However, this is just one aspect of the solution shared by many
other models too. The distinguishing features lie elsewhere. For instance, we
saw that in the scalar field model $\widetilde{M}_{pN}(r)=10^{6}\widetilde
{M}(r)$ [cf.Eq.(56)]. As emphasized earlier, such distinguishing features will
have to be determined only by the detailed analyses of data profiles of
potential, pressure and most importantly, the equation of state obtained
through the combined rotation curve and lensing measurements. The bottom line
is that the present solution depicts the halo as a \textit{mildly}
non-Newtonian model (pressure contribution is of the same order as energy
density) in contrast to the highly non-Newtonian scalar field model. Although
both the models share many properties including an anisotropic non-perfect
fluid equation of state, sharp differences appear in the computation of
$M_{pN}(r)$ and in the equation of state characterized by $\omega(r)$.
Observations to date do not seem yet conclusive enough (for a discussion, see
Faber \&\ Visser 2006) to tell us which one, if any, is closer to reality
among all the competitive halo models.

\textbf{ACKNOWLEDGMENTS}

We thank Arunava Bhadra for several enlightening discussions. KKN thanks Guzel
N. Kutdusova for her assistance at BSPU and SSPA where part of the work was
carried out.

\textbf{REFERENCES}

Agnese A.G. and La Camera M., 1995, Phys. Rev. D, 51, 2011

Alsing P.M., 1998, Am. J. Phys., 66, 779

Arbey A., Lesgourgues J. and Salati P., 2003, Phys. Rev. D, 68, 023511

Barnes D. G., Webster R. L., Schmidt R. W. and Hughes A.,1999, Mon. Not. R.
Astron. Soc., 309, 641

Begeman K. G., 1989, Astron. Astrophys., 223, 47

Bekenstein J. and Milgrom M., 1984, Astrophys. J., 286, 7

Bhadra A., Sarkar K. and Nandi K.K., 2007, Phys. Rev. D, 75,123004

Bhadra A., 2003, Phys. Rev. D, 67,103009

Bhadra A. and Sarkar K., 2005, Mod. Phys. Lett. A, 20,1831

Bharadwaj S. and Kar S., 2003, Phys. Rev. D, 68, 023516

Binney J.J. and Tremaine S., 1987, Galactic Dynamics (Princeton University Press)

Boonserm P., Cattoen C., Faber T., Visser M. and Weinfurtner S., 2005, Class.
Quant. Grav., 22, 1905

Boriello A. and Salucci P., 2001, Mon. Not. R. Astron. Soc., 323, 285

Cabral-Rosetti L.G., Matos T., Nu\~{n}ez D. and Sussman R.A., 2002, Class.
Quant. Grav.,19, 3603

Casadio R., Fabbri A. and Mazzacurati L., 2002, Phys. Rev. D, 65, 084040

Chandrasekhar S., 1983, Mathematical Theory of Black Holes (Oxford University Press)

Cheng Y.-C. N. and Krauss L. M., 1999, Astrophys. J., 514, 25

Creek S., Gregory R., Kanti P. and Mistry B., 2006, Class. Quant. Grav., 23, 6633

Dadhich N., Maartens R., Papadopoulos P. and Rezania V., 2000, Phys. Lett. B,
487, 1

Dadhich N., 2009, arXiv:gr-qc/0902.0205

Efstathiou G., Sutherland W. and Madox S. J., 1990, Nature, 348, 705

Einasto J., Kaasik A. and Saar E., 1974, Nature, 250, 309

Ellis J., Kaloper N., Olive K.A. and Yokoyama J., 1998, Phys. Rev. D, 59, 103503

Evans J.C., Alsing P.M., Giorgetti S. and Nandi K.K., 2001, Am. J. Phys., 69,1103

Faber T. and Visser M., 2006, Mon. Not. R. Astron. Soc., 372, 136

Fay S., 2004, Astron. Astrophys., 413, 799

Ford L.H. and Roman T.A., 1996, Phys. Rev. D, 53, 5496

Freeman K.C., 1970, Astrophys. J., 160, 881

Germani C. and Maartens R., 2001, Phys. Rev. D, 64, 124010

Guzm\'{a}n F. S. and Matos T., 2000, Class. Quant. Grav., 17, L9

Harko T. and Mak M. K., 2004, Phys. Rev. D, 69, 064020

Hochberg D. and Visser M., 1998, Phys. Rev. Lett., 81, 746

Kochanek C. S. and Schechter P. L., 2004, arXiv:astroph/0306040

Kochanek C. S., Falco E. E., Impey C., Lehar J., McLeod B., Rix H.-W., 2005,
CASTLE Survey Gravitational Lens Data Base, http://www.cfa.harvard.edu/glensdata/

Lake K., 2004, Phys. Rev. Lett., 92, 051101

Lidsey J.E., Matos T. and Ure\~{n}a-Lopez L.A., 2002, Phys. Rev. D, 66, 023514

Lobo F.S.N., 2008, in Classical and Quantum Gravity Research (Nova Sci. Pub.,
New York)

Lynden-Bell D., Katz J. and Bi\v{c}\'{a}k J., 2007, Phys. Rev. D, 75, 024040

Mak M. K. and Harko T., 2004, Phys. Rev. D, 70, 024010

Maoz E., 1994, Astrophys. J., 428, L5

Matos T., Guzm\'{a}n F. S. and Nu\~{n}ez D., 2000, Phys. Rev. D, 62, 061301

Metcalf R. B. and Silk J., 2007, Phys. Rev. Lett., 98, 071302

Mielke E.W. and Schunck F.E., 2002, Phys. Rev. D, 66, 23503

Misner C.W., Thorne K.S. and Wheeler J.A., 1973, Gravitation, (Freeman, San
Francisco, pp. 601-4)

Nandi K.K., Zhang Y.Z., Cai R.G. and Panchenko A., 2009, Phys. Rev. D, 79, 024011

Nandi K.K., Islam A. and Evans J., 1997, Phys. Rev. D, 55, 2497

Nandi K.K., Bhattacharjee B., Alam S.M.K. and Evans J., 1998, Phys. Rev. D,
57, 823

Nandi K.K., Zhang Y.Z. and Zakharov A.V., 2006, Phys. Rev. D, 74, 024020

Nandi K.K. and Islam A., 1995, Am. J. Phys., 63, 251

Nucamendi U., Salgado M. and Sudarsky D., 2001, Phys. Rev. D, 63, 125016

Oort J., 1930, Bull. Astron. Ins. Nether., V, 189

Ostriker P., Peebles P. J. E. and Yahill A., 1974, Astrophys. J. Lett., 193, L1

Peebles P.J. E., 2000, Phys. Rev. D, 62, 023502

Persic M., Salucci P. and Stel F., 1996, Mon. Not. R. Astron. Soc., 281, 27

Pope A. C. et al., 2004, Astrophys. J., 607, 655

Rahaman F., Kalam M., DeBenedictis A., Usmani A.A. and Ray S., 2008, Mon. Not.
R. Astron. Soc., 389, 27

Randall L. and Sundrum R., 1999a, Phys. Rev. Lett., 83, 3370

Randall L. and Sundrum R., 1999b, Phys. Rev. Lett., 83, 4690

Roberts M. S. and Rots A. H., 1973, Astron. Astrophys., 26, 483

Rubin V. C., Thonnard N. and Ford W. K., Jr., 1978, Astrophys. J. Lett., 225, L107

Rubin V. C., Roberts M. S. and Ford W. K., Jr., 1979, Astrophys. J., 230, 35

Sahni V., 2004, Lec. Notes Phys., 653, 141

Sanders R.H., 1984, Astron. Astrophys. 136, L21

Sanders R.H., 1986, Astron. Astrophys. 154, 135

Sarkar K. and Bhadra A., 2006, Class. Quant. Grav., 23, 6101

Shiromizu T., Maeda K. and Sasaki M., 2000, Phys. Rev. D, 62, 024012

Smith R. J., Blakeslee J. P., Lucey J. R. and Tonry J., 2005, Astrophys. J,
625, L103

Sofue Y. and Rubin V., 2001, Ann. Rev. Astron. \& Astrophys., 39, 137

Soleng H. H., 1995, Gen. Rel. Grav., 27, 367

Tegmark M. et al., 2004a, Phys. Rev. D, 69, 103501

Tegmark M. et al., 2004b, Astrophys. J., 606, 702

Trott C. M. and Webster R. L., 2002, Mon. Not. R. Astron. Soc., 334, 621

Visser M. and Wiltshire D. L., 2003, Phys. Rev. D, 67, 104004

Visser M., Kar S. and Dadhich N., 2003, Phys. Rev. Lett., 90, 201102

Viznyuk A. and Shtanov Y., 2007, Phys. Rev. D, 76, 064009

Weinberg N. N. and Kamionkowski M., 2002, Mon. Not. R. Astron. Soc., 337, 1269

Weinberg S., 1972, Gravitation \& Cosmology (John Wiley)

Zwicky F., 1933, Helvet. Phys. Acta, 6, 110

Zwicky F., 1937, Astrophys. J., 86, 217

\bigskip
\end{document}